\documentclass[aps,prx,groupaddress,twocolumn,floatfix]{revtex4-1}
\usepackage{units}
\usepackage{amsmath}
\usepackage{amssymb}
\usepackage{graphicx}
\usepackage{bm}
\usepackage{multirow,color,relsize,ulem,microtype}

\newcommand{\be}{\begin{equation}}
\newcommand{\ee}{\end{equation}}
\newcommand{\e}{\varepsilon}

\newcommand{\re}[1]{\text{Re}[#1]}
\newcommand{\im}[1]{\text{Im}[#1]}

\begin{document}

\title{Symmetry-protected zero-mode laser with a tunable spatial profile}

\author{Li Ge}
\email{li.ge@csi.cuny.edu}
\affiliation{\textls[-18]{Department of Engineering Science and Physics, College of Staten Island, CUNY, Staten Island, NY 10314, USA}}
\affiliation{The Graduate Center, CUNY, New York, NY 10016, USA}

\date{\today}

\begin{abstract}
We propose to utilize symmetry-protected zero modes of a photonic lattice to realize a single-mode, fixed-frequency, and spatially tunable laser. These properties are the consequence of the underlying non-Hermitian particle-hole symmetry, with which the energy spectrum satisfies $\e_m=-\e_n^*$. Unlike in the Hermitian case, the symmetric phase of particle-hole symmetry is no longer restricted to $\e=0$ but extends along the imaginary-$\e$ axis, which is set by the single-cavity frequency and symmetry-protected against position and coupling disorder of the photonic lattice. By selectively pumping different cavities in the photonic lattice, we control the spontaneous symmetry restoration process, which provides a convenient method to tune the spatial profile of the laser without changing its frequency.
\end{abstract}

\maketitle

\section{Introduction}

Originally proposed as a model for neutrinos, Majorana zero modes have attracted considerable interest in the past decade \cite{Hasan,Qi,Alicea,Beenakker}. Their topological and non-Abelian properties have stimulated enormous research efforts in robust topological quantum computation, and condensed matter systems such as the 5/2-fractional quantum Hall liquid and semiconductor nanowires have been suggested as potential platforms to verify the existence of these exotic states \cite{Sarma_RMP,Sarma_QI,Alicea_PRX}. Meanwhile, robust zero modes also exist in photoinc systems \cite{Lu}, which can be both symmetry protected and topologically protected \cite{Malzard}. Although the effective Hamiltonians $\tilde{H}$ of these photonic systems are necessarily non-Hermitian due to the lack of photon number conservation, an intimate connection exists between them and the corresponding off-shell scattering matrices $S(\e)$ at a complex-valued energy $\e$ \cite{EP_CMT}.

It was shown that the Bogolyubov-de Gennes equation in the Majorana basis leads to a scattering matrix satisfying $S^*(\e)=S(-\e)$ \cite{Pikulin}, where ``$*$" denotes the complex conjugate. This relation indicates the existence of zero modes, now defined by the poles of the $S$ matrix on the imaginary-$\e$ axis. They are the consequence of the underlying particle-hole symmetry \cite{Ryu} (also known as charge conjugation symmetry): the Hamiltonian anticommutes with an anti-unitary operator $\cal CT$, where $\cal C$ is unitary and $\cal T$ is the time reversal operator. The same symmetry also leads to zero modes in a proposed photonic structure with asymmetric couplings \cite{Malzard}.

In fact, zero modes in photonics have been observed in various lattice systems, where identical cavities or waveguides are coupled by evanescent waves. These experimental demonstrations include, for example, the zeroth Landau level in a strained honeycomb lattice \cite{Rechtsman}, the flat band in Lieb lattices \cite{Mukherjee,Vicencio,Baboux}, and a mid-gap defect state in a SSH chain \cite{Schomerus}. The existence of zero modes in these experiments can be understood from the chiral symmetry (also known as sublattice symmetry) of the system in the Hermitian limit $\tilde{H}\rightarrow H$, where $H$ anticommutes with a unitary operator.

In the Hermitian case, both chiral symmetry and particle-hole symmetry lead to a (real-valued) symmetric energy spectrum about $\e=0$, and zero modes are typically excited states of the system and difficult to observe (with exceptions, for example, in $p$+i$p$ superconductors \cite{Sarma_RMP}). In the non-Hermitian case, however, the energy eigenvalues now become complex in general, and their imaginary parts indicate the overall decay rates of photons in the corresponding modes due to scattering, absorption and other processes. As a result, chiral symmetry and particle-hole symmetry now have \textit{different} consequences: the chiral symmetry retains $\e_m=-\e_n$, whereas the particle-hole symmetry gives rise to $\e_m=-\e_n^*$ \cite{Pikulin}; zero modes form if $m=n$ in these relations, i.e., they feature $\e_n=0$ and $\re{\e_n}=0$ due to these two symmetries, respectively.

The imaginary part of $\e$ changes the observability of these non-Hermitian zero modes. Since the dynamics of an energy eigenstate $\Psi_n(\bm{x},t)$ is given by $e^{-i\e_n t}$ ($\hbar\equiv1$), a non-Hermitian system with chiral symmetry is typically unstable due to energy eigenvalues in the upper complex plane ($\im{\e}>0$). An exception occurs if all energy eigenvalues are real, which is possible in a non-Hermitian system if it also has parity-time ($\cal PT$) symmetry  \cite{Bender1,El-Ganainy_OL06,Moiseyev,Makris_prl08,Longhi,RC,Microwave,Regensburger,CPALaser,PTConservation,Robin,Ge_PRX,Feng,Hodaei,Yang,PT_RMP}. The excitation of a zero mode in this case, nevertheless, faces a similar challenge as in a Hermitian system, and a resonant drive at the zero energy is typically required \cite{Rechtsman,Mukherjee,Vicencio}. For the particle-hole symmetry, however, the entire non-Hermitian spectrum can reside in the lower half of the complex plane ($\im{\e}<0$) due to the aforementioned relation $\e_m=-\e_n^*$. We can then selectively pump a zero mode $\Psi_n(\bm{x},t)$ and raise the corresponding $\e_n(=-\e_n^*)$ to the real axis before any other modes, at which the decay rate of $\Psi_n(\bm{x},t)$ is compensated exactly by the pump and $\Psi_n(\bm{x},t)$ undergoes a sustained oscillation.

We note that this excitation strategy is exactly to bring a mode to its lasing threshold, where $\e_m$ corresponds to a pole of the scattering matrix (see, for example, Refs.~\cite{EP_CMT,SPASALT}). While it adds a new tool to the study of zero modes using photonic lattices, this approach falls short of providing new inspirations for a better design and control of laser properties: the ``zero energy" is in fact the resonant frequency of a single cavity ($\omega_0$), and a zero-mode photonic lattice laser provides the same lasing frequency as a single-cavity laser.

In this work we present an exciting perspective against this view, by exploring the spatial degrees of freedom offered by zero modes. More specifically, by pumping different cavities in a photonic lattice, we are able to tune the spatial profile of the lasing zero mode, while its frequency is pinned at the zero energy and protected against position and coupling disorder of the photonic lattice. The result is a single-mode, fixed-frequency, and spatially tunable laser, which to the best of our knowledge, is the first practical application of zero modes in photonics. In fact, we do not even require that the system has zero modes before the pump is applied. As we shall see, they can be created automatically by a spontaneous restoration process of the underlying non-Hermitian particle-hole symmetry. We note that ``single-mode" here refers to a laser with a single lasing frequency, and ``zero-mode" means that this lasing frequency is at $\e=0$.

Below we first discuss briefly the spontaneous restoration of chiral symmetry and particle-hole symmetry in non-Hermitian systems. We then focus on the latter and exemplify it in a rectangular photonic lattice, where we explain why selective pumping leads to non-Hermitian particle-hole symmetry when the system has two sublattices. Furthermore, we analyze the aforementioned zero-mode laser and show its different spatial profiles and unique phase distributions when using different pump profiles. We also study disorders that lift non-Hermitian particle-hole symmetry in such a laser, which suggests that its frequency accuracy is comparable or even better than a single-cavity laser. Finally, we discuss non-Hermitian particle-hole symmetry in previous demonstrated $\cal PT$-symmetric photonic-molecule lasers \cite{Hodaei,Yang} and more complicated photonic structures.

\section{Spontaneous symmetry restoration}

\subsection{Chiral symmetry}

Chiral symmetry applies to a system consisting of two sublattices $A$ and $B$, where couplings only take place between two cavities belonging to different sublattices. A non-Hermitian effective Hamiltonian with chiral symmetry can be brought to the following form:
\be
\tilde{H} =
\begin{pmatrix}
0 & T_1 \\
T_2 & 0
\end{pmatrix},\label{eq:H_chiral*}
\ee
where $T_1\neq T_2^\dagger$ and they are $K\times L$ in size. ``$\dagger$" denotes the Hermitian conjugate as usual, and $K,L$ are the sizes of the two sublattices. It is straightforward to verify that $\tilde{H}$ given by Eq.~(\ref{eq:H_chiral*}) anticommutes with a diagonal operator ${\cal C}$ consisting of $K$ 1's followed by $L$ $-1$'s. $\cal C$ can be written as ${\mathbf 1}_L\otimes \sigma_z$ when $K=L$, where ${\mathbf 1}_L$ is the identity matrix of rank $L$ and $\sigma_z$ is the third Pauli matrix.

\begin{figure}[t]
\includegraphics[clip,width=\linewidth]{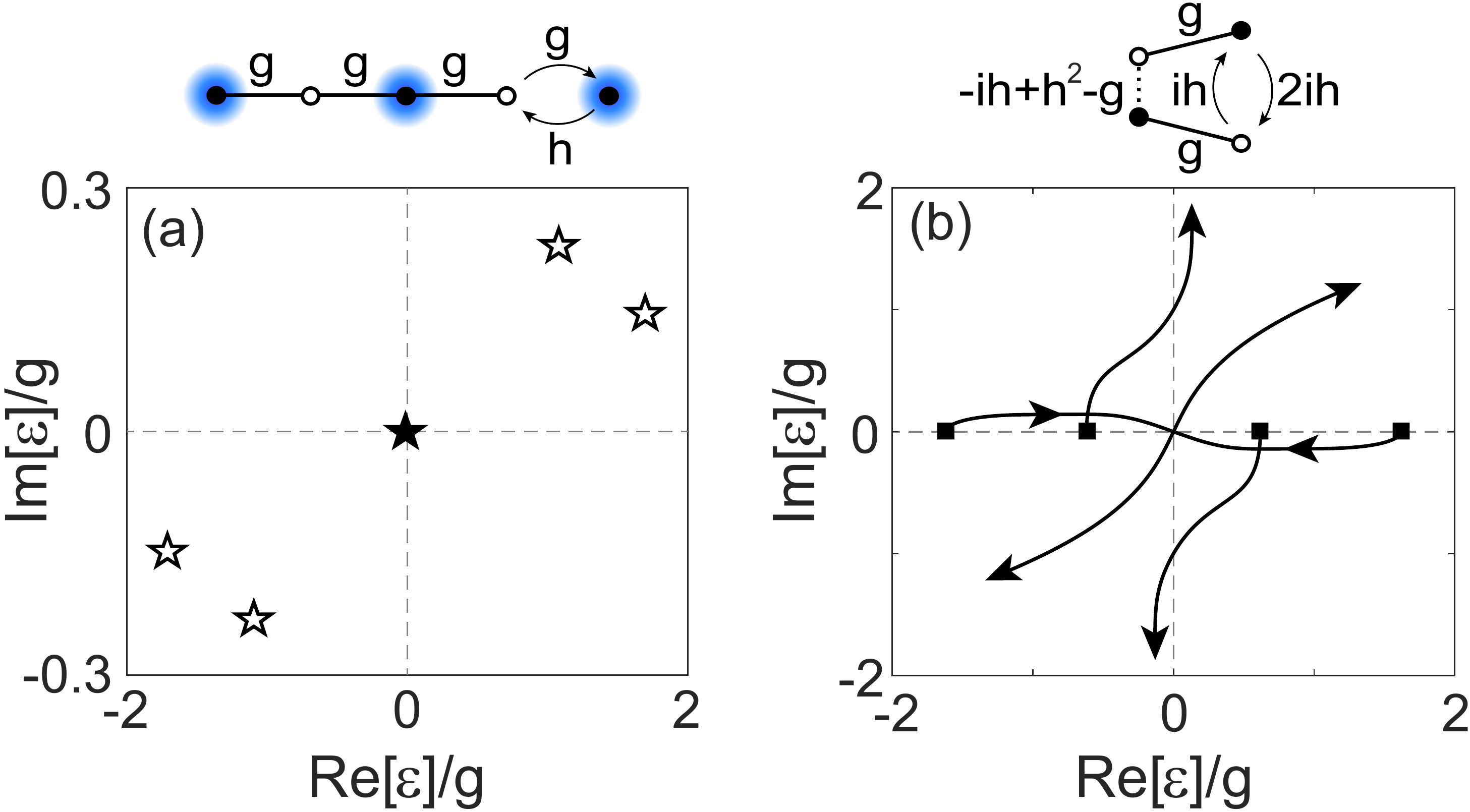}
\caption{(a) Complex energy spectrum (stars) of a one-dimensional (1D) non-Hermitian lattice with chiral symmetry. The effective Hamiltonian $\tilde{H}$ is given by the general expression (\ref{eq:H_chiral*}) with the couplings shown in the schematic, where $h=(1+i)g$. Filled and open circles indicate sublattices $A$ and $B$, and the shades on the lattice represent the zero-energy dark state.  (b) Trajectories of $\e_n$ (solid lines) as $h$ increases from 0 (squares) to $1.5g$. The lattice and couplings are shown in the schematic. }\label{fig:chiral*}
\end{figure}

$\tilde{H}$ has an eigenvalue spectrum that satisfies the aforementioned property $\e_m = -\e_n$ (see the toy models in Fig.~\ref{fig:chiral*}), which can be understood by noticing that if $\Psi_n$ and $\e_n$ are one eigenstate and the corresponding eigenvalue of $\tilde{H}$, then $\Psi_m={\cal C}\Psi_n$ is another eigenstate of $\tilde{H}$ with $\e_m = -\e_n$. For the special case that $m=n$, we have zero mode(s) $\e_n=0$ and the corresponding eigenstate(s) satisfies $\Psi_{n}={\cal C}\Psi_{n}e^{i\theta_{n}}={\cal C}^2\Psi_{n}e^{2i\theta_{n}}$. Given that the aforementioned operator $\cal C$ satisfies ${\cal C}^2=1$, $\theta_{n}$ can only be 0 or $\pi$ and as a result, $\Psi_{n}$ vanishes on one of the sublattices, which can be referred to as a dark state. We note that the Lieb's theorem \cite{LiebThm} still applies in the non-Hermitian case, i.e., there are at least $|K-L|$ symmetry-protected dark states on sublattice $A(B)$ when $K>L$ $(L<K)$.

Because the dark states possess the same (chiral) symmetry as the system itself, they define the symmetric phase of the effective Hamiltonian, whereas the finite-energy modes satisfying $\Psi_m={\cal C}\Psi_n (m\neq n)$ represent the broken-symmetry phase. Spontaneous restoration of chiral symmetry is possible, i.e., two finite-energy eigenvalues become zero simultaneously. This situation, however, takes place only at singular point(s) in the parameter space, where the matrix rank of $T_1$ or $T_2$ becomes lower than min($K,L$). In Fig.~\ref{fig:chiral*}(b) we show one example for a toy model consisting of four cavities, and two modes coalesce at $\e=0$ when the parameter $h$ is equal to $g$, where the rank of $T_2$ reduces to 1 and becomes smaller than $K=L=2$.

\subsection{Particle-hole symmetry}

\begin{figure}[b]
\includegraphics[clip,width=\linewidth]{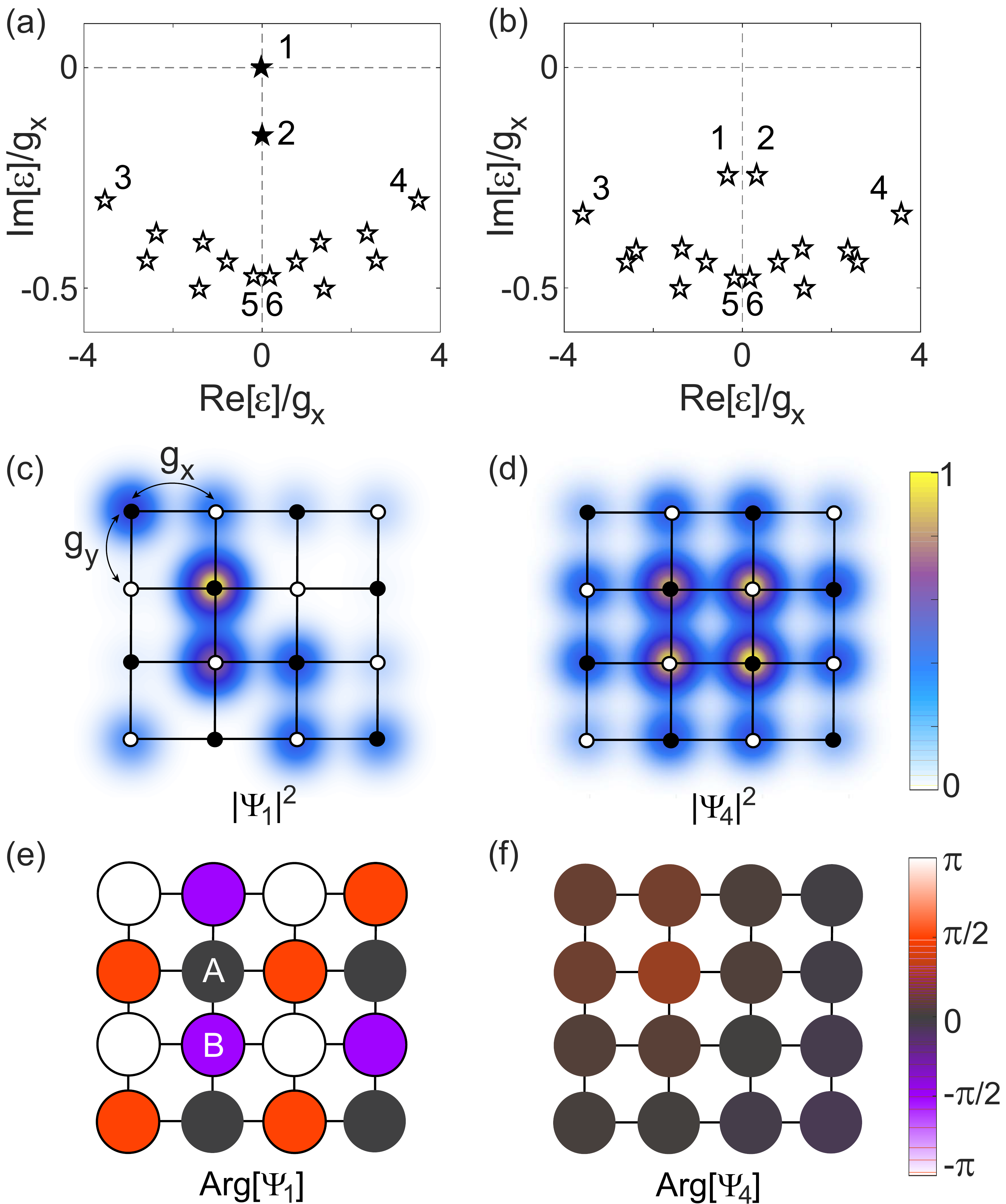}
\caption{(a) Complex energy spectrum of a rectangular lattice with non-Hermitian particle-hole symmetry. The system is pumped at the $A$ cavity marked in (e) with pump strength $\gamma_A=2.0g_x$. The couplings are shown in (c) and satisfy $g_y=1.25g_x\in{\mathbb R}$, and the cavity decay rate is given by $\kappa_0=0.5g_x$. (b) Same as (a) but with a lower pump strength ($\gamma_A=1.5g_x$) and no zero modes.
(c),(e) Schematics showing the spatial profile and phase distribution of mode 1 in (a). Gaussian packets are used to represent $|\Psi_1|^2$ in (c). (d), (f) Same as (c), (e) but for mode 4 in (a).
}\label{fig:ph*}
\end{figure}

Similar to the discussion above, the eigenstates of a system with non-Hermitian particle-hole symmetry can also exist in two phases, i.e., the symmetric phase where $\Psi_n = e^{i\theta_n}{\cal CT}\Psi_n$, $\e_n=-\e_n^*$ and the broken-symmetry phase where $\Psi_m = {\cal CT}\Psi_n$, $\e_m=-\e_n^*\,(m\neq n)$ [see Fig.~\ref{fig:ph*}(a)]. The major differences between the spontaneous symmetry restoration of these two symmetries are two-fold.
First, the symmetric phase here is no longer restricted to where $\e=0$; it now extends in the parameter space where $\e_n$ is imaginary. As such, spontaneous restoration of non-Hermitian particle-hole symmetry is relatively easy to achieve, as we exemplify below using a two-dimensional photonic lattice.

In addition, $\Psi_n$ in the symmetric phase are no longer dark states that have non-zero intensity only on one sublattice [see Fig.~\ref{fig:ph*}(c)].
This is because $\Psi_n=e^{i\theta_n}{\cal CT}(e^{i\theta_n}{\cal CT}\Psi_n)=({\cal CT})^2\Psi_n$ holds automatically [since $({\cal CT})^2=1$]; this arbitrary angle $\theta_n$ versus its value of 0 or $\pi$ in the chiral symmetric phase is analogous to the additional phase angle of the wave function introduced by exchanging two anyons versus bosons or fermions \cite{Wilczek}. For a system with two sublattices $A$, $B$ and the aforementioned diagonal operator $\cal C$, the only requirement on $\Psi_n\equiv(\begin{smallmatrix}\Psi_A\\\Psi_B\end{smallmatrix})$ in the symmetric phase is that $2\text{Arg}(\Psi_A)_p$, $2\text{Arg}(\Psi_B)_q$ are independent of the lattice positions $p,q$ and they differ by $\pi$. In other words, $\Psi_A$ can be made real while $\Psi_B$ imaginary by choosing a proper global phase of $\Psi_n$ [see Fig.~\ref{fig:ph*}(e)]. This unique phase distribution distinguishes the wave function of a zero mode from that of a finite-energy mode in the broken phase of particle-hole symmetry [see Fig.~\ref{fig:ph*}(f)].

\section{Zero-mode laser}

\subsection{Model system}
\label{sec:model}
In this section we first explain why the photonic lattice with locally pumped and otherwise identical cavities shown in Fig.~\ref{fig:ph*} can be described by an effective Hamiltonian with particle-hole symmetry. We start by noting that many experimentally realizable photonic lattices consist of two sublattices, including the square lattice, honeycomb lattice \cite{Rechtsman}, Lieb lattice \cite{Mukherjee,Vicencio,Baboux} and so on. The couplings between the two sublattices can be represented by a Hermitian matrix $H$ that satisfies chiral symmetry:
\be
H =
\begin{pmatrix}
0 & T \\
T^\dagger & 0
\end{pmatrix}.
\ee
The effective Hamiltonian of the non-Hermitian photonic lattice is completed by including the cavity decay rate $\kappa_0$ and the pump strength $\gamma_p$:
\be
\tilde{H} = H+i(\gamma_p-\kappa_0)\delta_{pq}\equiv H + iH_1~(\gamma_p,\kappa_0\in{\mathbb R}), \label{eq:H_ph*}
\ee
where $\delta_{pq}$ is the Kronecker delta. The real diagonal matrix $H_1$ commutes with the unitary operator $\cal C$ specified before, and hence $iH_1$ satisfies particle-hole symmetry but not chiral symmetry. Now if the couplings in $T$ are real (in the case of evanescent wave coupling, for example), then $H$ has particle-hole symmetry as well and so does $\tilde{H}$.

If all $\gamma_p$ are the same (denoted by $\gamma$), then the spectrum of $\tilde{H}$ is simply shifted from that of $H$ by $\gamma-\kappa_0$ along the imaginary-$\e$ axis. Apparently varying $\gamma$ does not lead to spontaneous symmetry restoration or breaking of the eigenstates, nor does it favor a zero mode (if exists) to be the lasing mode.
Therefore, we are more interested in the case that not all $\gamma_p$ are the same, i.e., we require a selective pump configuration where $\gamma_p$ vary from cavity to cavity. Such a selective pumping scheme has been employed in lasers to explore phenomena such as laser self-termination \cite{EP7,EP_CMT,EP_exp,Nonl_PT} and chaos-assisted tunneling \cite{Narimanov}. 

In Fig.~\ref{fig:ph*} we have shown one example of a rectangular lattice with four cavities on each side and nearest neighbor couplings $g_{x,y}$ in the horizontal and vertical directions. Before the pump is applied, the system (and its eigenstates) is separable along the two directions, i.e., $\Psi_n(x,y) = \psi(x)\phi(y)$, and the 1D sublattices that determine $\psi(x),\phi(y)$ satisfy non-Hermitian particle-hole symmetry as well. As a result, a pair of $\psi_+(x)={\cal CT}\psi_-(x)$ and a pair of $\phi_+(y)={\cal CT}\phi_-(y)$ form a quartet that have the same intensity pattern $|\Psi(x,y)|^2$, such as mode 1--4 in Figs.~\ref{fig:ph*}(a) and (b) before the pump is applied. Note that due to the different values of $g_x,g_y$, all modes of the system are in the broken phase of particle-hole symmetry when the pump is weak, with all energy eigenvalues away from the imaginary axis [see Fig.~\ref{fig:ph*}(b), for example].

\begin{figure}[b]
\includegraphics[clip,width=\linewidth]{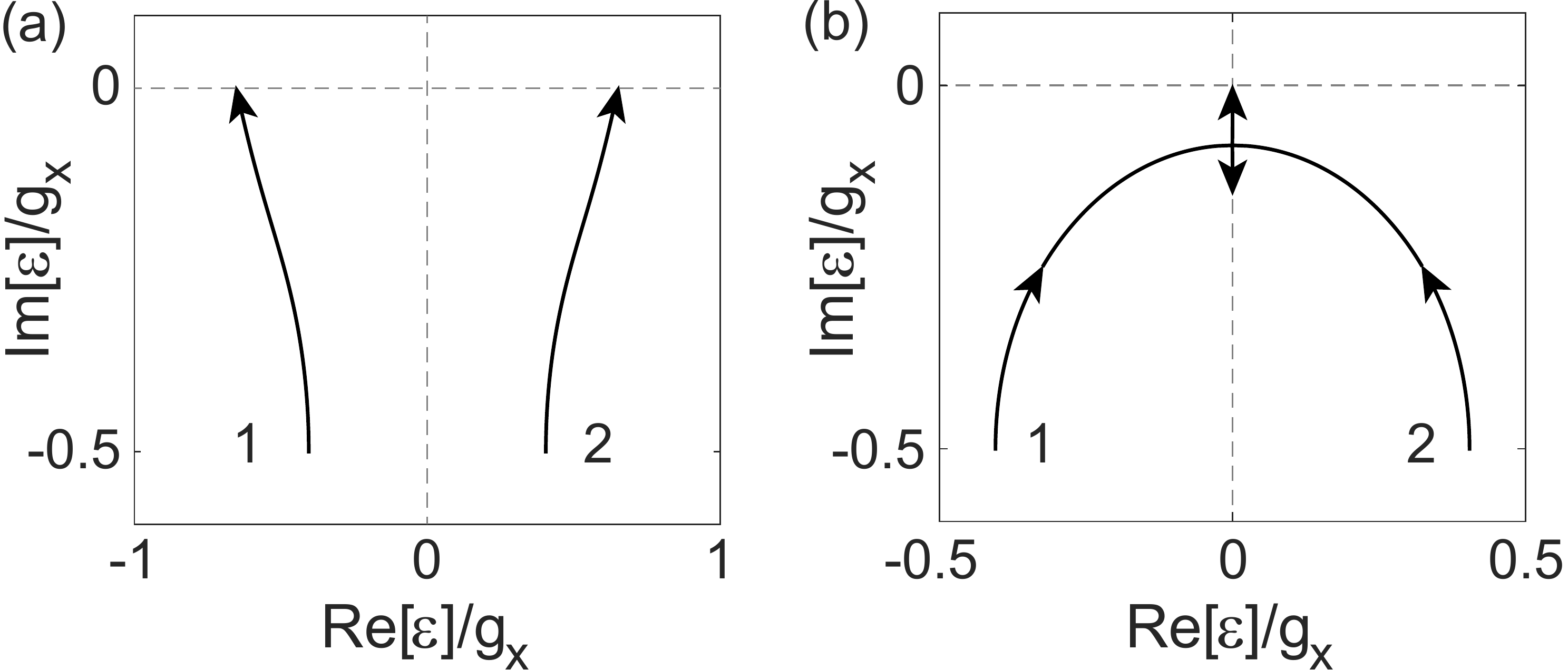}
\caption{Spontaneous restoration of non-Hermitian particle-hole symmetry (b) and the lack of it (a) for mode  1 and 2 in Fig.~\ref{fig:ph*}(a) as the pump strength increases. The laser threshold is reached at $\gamma_A=\gamma_B\approx1.40g_x$ in (a) and $\gamma_A\approx2.02g_x$ in (b). }\label{fig:ph*_breaking}
\end{figure}

We have chosen the cavity decay rate $\kappa_0=g_x/2$, and we find that no spontaneous symmetry restoration occurs if we pump a pair of $A$ and $B$ cavities near the center [marked in Fig.~\ref{fig:ph*}(e)] to the lasing threshold of the system, defined by $\max(\im{\e_n})=0$ [see Fig.~\ref{fig:ph*_breaking}(a)]. However, the situation changes when we just pump the marked $A$ cavity: mode 1 and 2 coalesce at an exceptional point \cite{EP1,EP2,EPMVB,EP3,EP4,EP5,EP6,EP8} on the imaginary axis and move along it as the pump strength $\gamma_A$ is increased [see Fig.~\ref{fig:ph*_breaking}(b)]. One of them reaches the real-$\e$ axis before all other modes and becomes the lasing mode of the system.
The experimental signatures of these two different scenarios can be easily distinguished. In the case with restored non-Hermitian particle-hole symmetry, there is only one lasing mode and its frequency is given by the single-cavity frequency ($\re{\e}=0$). In the case with broken particle-hole symmetry, two eigenstates emerge as the lasing modes if the gain is inhomogeneously broadened (i.e., cross saturation is weak when the frequencies of these modes are different), both detuned from the single-cavity frequency.

We note that although these properties are similar to a $\cal PT$-symmetric laser \cite{CPALaser,Longhi,Feng,Hodaei,Yang}, our system clearly lacks $\cal PT$ symmetry with the aforementioned pump configurations: there does not exist a parity operation that exchanges gain and loss in the system, and in addition, the spectrum satisfy $\e_m=-\e_n^*$ instead of $\e_m=\e_n^*$ [see Figs.~\ref{fig:ph*}(a) and (b)]. In fact, the two different lasing scenarios exist in a $\cal PT$-symmetric laser exactly because it also has particle-hole symmetry, which we will elucidate in Sec.~\ref{sec:PT}.

\subsection{Tunable spatial profile at fixed frequency}

In this subsection we show that by pumping different cavities in the photonic lattice, the spontaneous restoration process of particle-hole symmetry is modified, which provides a convenient approach to change the spatial profile of the laser without affecting its frequency, pinned at the zero energy and given by the single-cavity frequency.

\begin{figure}[b]
\includegraphics[clip,width=\linewidth]{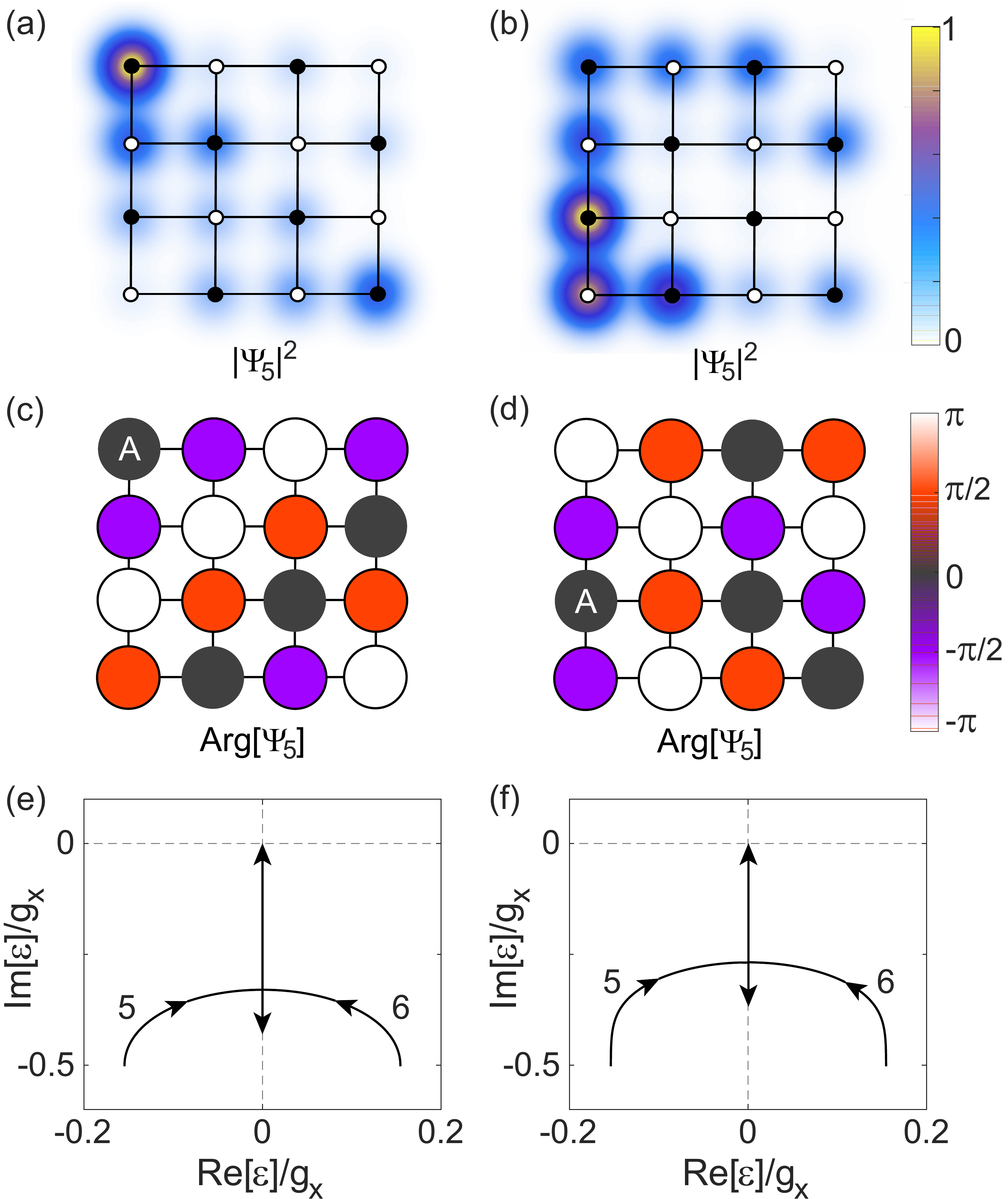}
\caption{Tunable spatial profile of a zero-mode photonic lattice laser. (a),(c) Spatial profile and phase distribution of mode 5 in Fig.~\ref{fig:ph*}(a) when pumping the $A$ cavity at the upper left corner of the rectangular lattice. (b),(d) The same but when pumping the $A$ cavity marked in (d). }\label{fig:ph*_profile}
\end{figure}

We have shown that by pumping the $A$ cavity near the center of the rectangular lattice, the lasing zero mode is created by the spontaneous restoration of the particle-hole symmetry for mode 1 and 2 shown in Figs.~\ref{fig:ph*}(a) and \ref{fig:ph*_breaking}(b). Now if we pump the $A$ cavity at the upper left corner of the rectangular lattice instead, modes 5 and 6 in Fig.~\ref{fig:ph*}(a) undergo spontaneous symmetry restoration instead [see Fig.~\ref{fig:ph*_profile}(e)], leading to a different spatial profile of the laser at threshold [see Fig.~\ref{fig:ph*_profile}(a)]. In fact, a localized pump profile creates a strong mixing of modes in the passive system (i.e., before the pump is applied).
As a result, the same pair of passive modes can give rise to different zero modes at laser threshold, when we vary the pump profile. One example is shown in Fig.~\ref{fig:ph*_profile}(b), where we pump the lower $A$ cavity in the leftmost column and the lasing mode is again due to the spontaneous symmetry restoration of mode 5 and 6 [see Fig.~\ref{fig:ph*_profile}(f)]. Besides the spatial profile, the phase distribution of the wave function also changes, as we show in Figs.~\ref{fig:ph*_profile}(c) and (d).

We have verified that these spatial properties change when pumping each $A$ cavity in our example, and each time the lasing takes place in a zero mode with unchanged lasing frequency. The same observations hold when pumping more than one cavity. For example, they exist in 22 out of 28 two-cavity pumping configurations on sublattice $A$ with equal strength and also when pumping the entire sublattice $A$ uniformly [see Figs.~\ref{fig:ph*_breaking2}(a),(b) in Sec.~\ref{sec:PT}]. In addition, since sublattices $A$ and $B$ are mapped onto each other with a parity operation about the central line either in the horizontal or vertical direction, each pumping configuration on sublattice $A$ has a corresponding configuration on sublattice $B$, which then doubles the available spatial profiles of this zero-mode photonic lattice laser. Without further enumerating all multi-cavity pumping configurations on sublattice $A$ and (or) $B$, we already see that this approach leads to a very versatile laser with a tunable spatial profile \textit{and} a fixed lasing frequency, thanks to the zero modes protected by the non-Hermitian particle-hole symmetry.

\begin{figure}[b]
\includegraphics[clip,width=\linewidth]{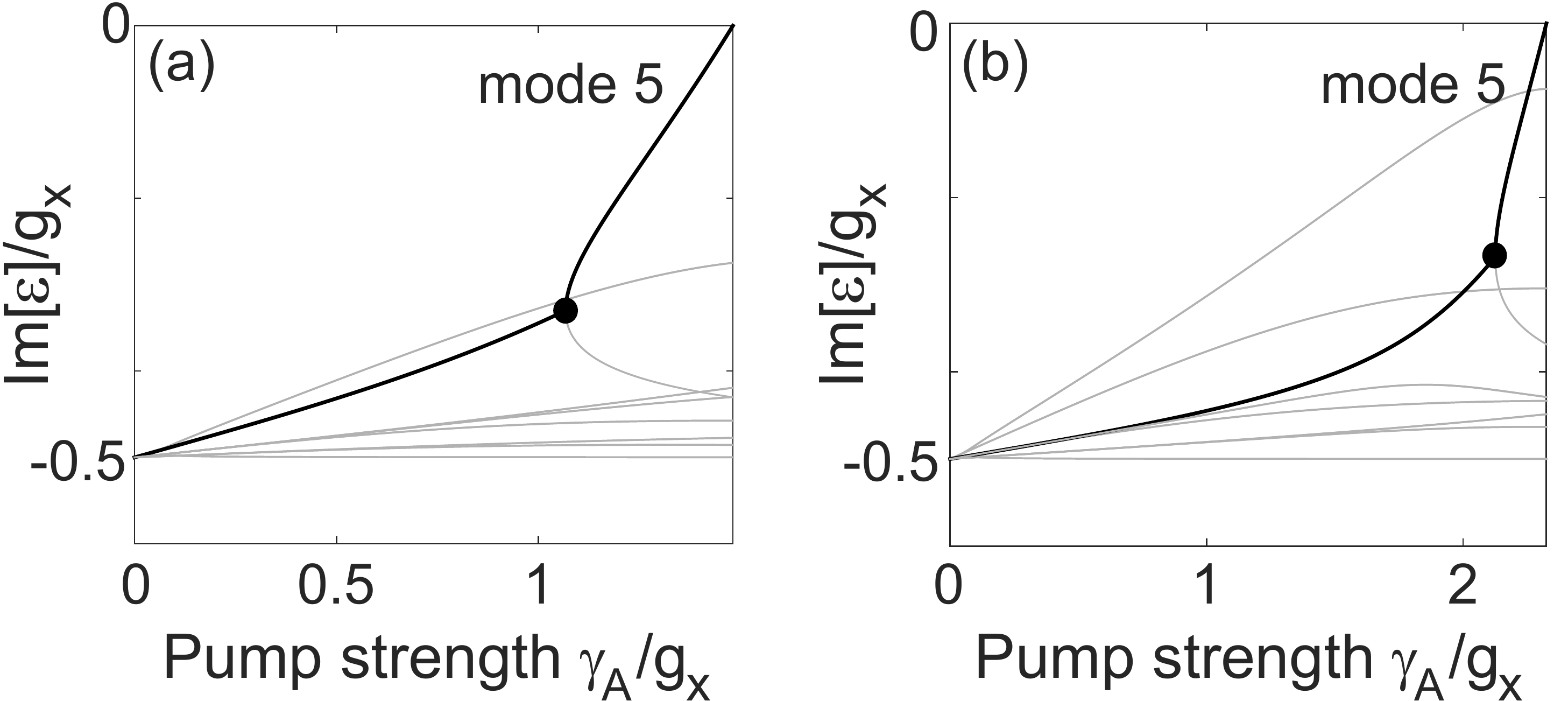}
\caption{Boosted gain of the lasing zero mode (thick black line) after the exceptional point (filled dot). (a) and (b) correspond to the two different pumping configurations in Fig.~\ref{fig:ph*_profile}. Thin grey lines show the other 15 supermodes originated from the same single-cavity frequency.}\label{fig:ph*_imag}
\end{figure}

It is worth noting that the lasing zero mode in each pump configuration here goes through an exceptional point when the spontaneous symmetry restoration takes place. This process leads to an enhanced gain of this mode, which in turn leads to a good single-mode performance: the imaginary part of this zero mode is significantly boosted after the exceptional point (see Fig.~\ref{fig:ph*_imag}), which pushes it to be the first mode to reach the laser threshold (again defined by $\im{\e}=0$), even way ahead of others in some cases [see Fig.~\ref{fig:ph*_imag}(a)].

To discuss the single-mode performance of our photonic lattice laser, it is also important to note that so far we have only discussed the ``supermodes" that originate from the same single-cavity frequency $\omega_0$, assumed to be at or closest to the center frequency of the gain curve where the effective pump strength $\gamma$ is the strongest. There are other groups of supermodes formed by different $\omega_0$'s, but these modes have a lower $\gamma$, and by noting the steep increase of gain ($\im{\e_n}$) with $\gamma$ in Fig.~\ref{fig:ph*_imag} for the lasing zero mode after the exceptional point, we know that these other competing groups of supermodes will have a noticeably higher lasing threshold, a concept that has been demonstrated in $\cal PT$-symmetric lasers \cite{Hodaei}.

\subsection{Effect of disorder}

In the discussions above we have shown one property of the photonic lattice laser that is protected by non-Hermitian particle-hole symmetry: the frequency of the lasing zero mode is independent of how we pump the photonic lattice. As we have explained above, the zero energy corresponds to the center frequency of the gain curve, hence this property is a special (symmetry-protected) case of ``line pulling" (see Ref.~\cite{Tureci_PRA}, for example), where the passive cavity frequencies (now of the supermodes) are pulled towards the gain center as the pump strength increases. 

The symmetry protection of the lasing frequency also applies to position and coupling disorder of the cavities in the photonic lattice. As long as these disorders do not change the nature of the couplings (i.e., by evanescent waves and real-valued), non-Hermitian particle-hole symmetry of the effective Hamiltonian stays intact, even when the couplings become asymmetric, i.e., $\tilde{H}_{pq}\neq\tilde{H}_{qp}(p\neq q)$. Therefore, the lasing frequency of a zero mode is still pinned at $\e=0$. 

\begin{figure}[b]
\includegraphics[clip,width=\linewidth]{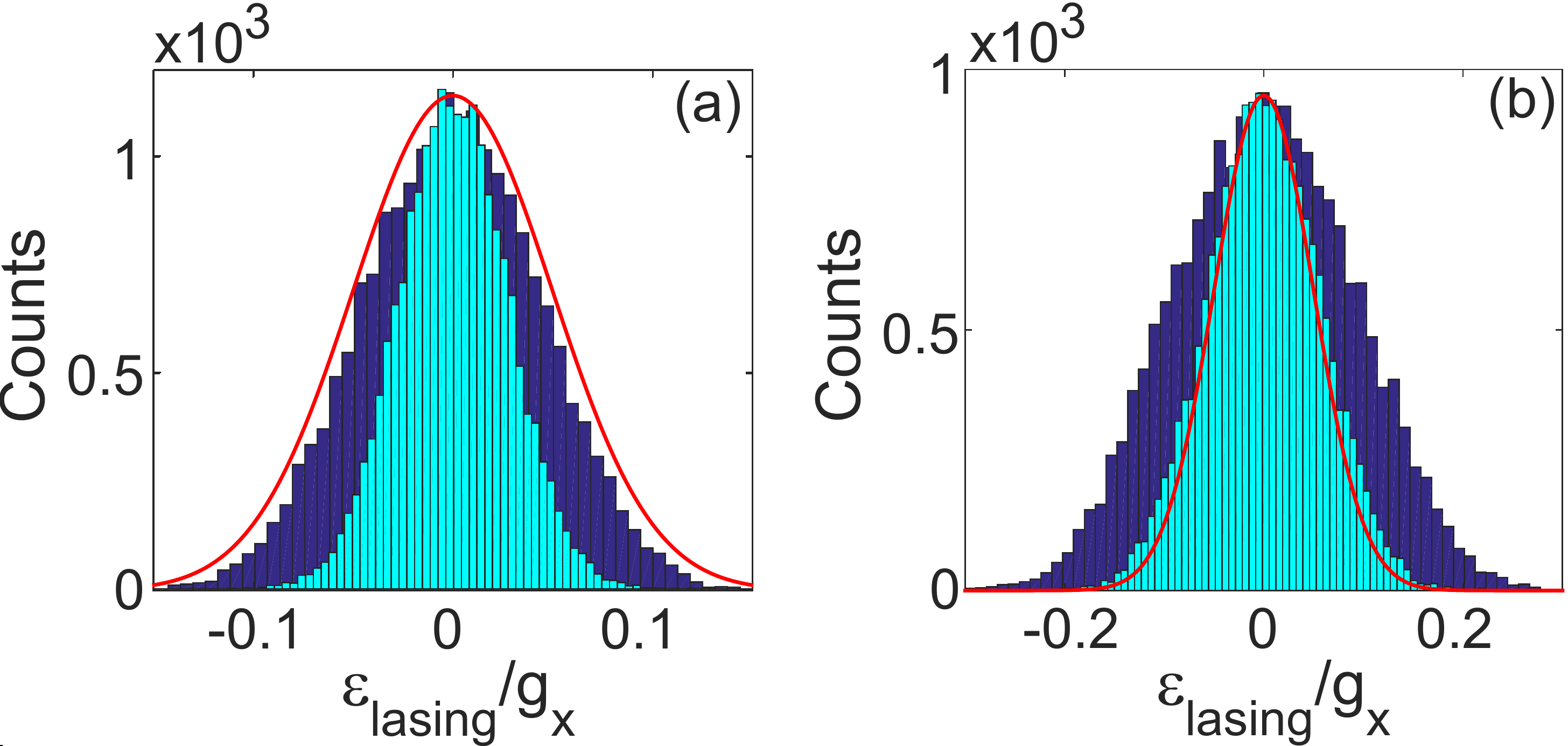}
\caption{Effect of disorder that break particle-hole symmetry. (a) Histogram of the lasing frequency for 20,000 realizations of $\omega_0$ disorder (dark blue bars) and NNN coupling disorder (light blue bars), both with a standard deviation of $g_x/20$ (red solid curve). The pump configuration is the same as in Fig.~\ref{fig:ph*_imag}(a). (b) Same as (a) but for the pump configuration in Fig.~\ref{fig:ph*_imag}(b). }\label{fig:ph*_disorder}
\end{figure}

Similar to chiral symmetry in the Hermitian case, particle-hole symmetry here breaks down when the single-cavity frequency varies from cavity to cavity, or when there are higher-order couplings between cavities in the same sublattice. For pump configurations where the gain of the lasing zero mode is much higher than the rest [see Fig.~\ref{fig:ph*_imag}(a)], we find that the variation of the lasing frequency $\e_\text{lasing}$ from zero energy is in fact smaller than that of the disorder. One example is shown in Fig.~\ref{fig:ph*_disorder}(a), where 20,000 realizations of Gaussian-distributed $\omega_0$'s are examined. The standard deviation of $\e_\text{lasing}$ (denoted by $\sigma_\text{lasing}$) is about 90\% of that of a weak $\omega_0$ in this case. 
The variation of the lasing frequency is even smaller for the disorder of next-nearest-neighbor (NNN) couplings, which take place diagonally between two neighboring cavities in the same sublattice. For the same pump configuration considered above, $\sigma_\text{lasing}$ is about 55\% of the standard deviation of NNN couplings, with the latter again modeled as a weak Gaussian disorder. When there are competing finite-energy modes with close thresholds as in Fig.~\ref{fig:ph*_imag}(b), these finite-energy modes only have a small chance of becoming the first lasing mode when the disorder is weak [e.g., about 0.01\% for $\omega_0$ disorder and 1\% for NNN coupling disorder in Fig.~\ref{fig:ph*_disorder}(b)], even though $\sigma_\text{lasing}$ becomes slightly larger than the standard deviation of the disorder [about 1.7 and 1.1 times for the two types of disorder in Fig.~\ref{fig:ph*_disorder}(b)]. Therefore, we can conclude that the frequency accuracy of our zero-mode photonic lattice laser is comparable or even higher than that of a single-cavity laser in the presence of fabrication disorder, which is a desirable property towards practical applications.

\subsection{Relation to $\cal PT$-symmetric lasers}
\label{sec:PT}
Non-Hermitian particle-hole symmetry is a general property of coupled systems with two sublattices in the presence of gain and loss, which we have shown using Eq.~(\ref{eq:H_ph*}). As mentioned at the end of Sec.~\ref{sec:model}, particle-hole symmetry in fact exists in several previous demonstrations of $\cal PT$ lasers. Take the photonic molecule (PM) lasers \cite{Hodaei,EP_exp,Yang} for example. They consist of two coupled identical microring, microdisk or microtoroid cavities, and their effective Hamiltonian can be written as a $2\times2$ matrix:
\be
\bar{H}=
\begin{pmatrix}
i\gamma_1 & g \\
g & -i\kappa_2
\end{pmatrix}.\label{eq:H_PT}
\ee
The parity operator ${\cal P}$ exchanges these two cavities and takes the form of the first Pauli matrix $\sigma_x$. 
If the net gain in one cavity equals the loss in the other ($\gamma_1=\kappa_2$), the system is $\cal PT$-symmetric and satisfies $({\cal PT}) \bar{H} ({\cal PT}) = \bar{H}$, with its energy eigenvalues either being real ($\cal PT$-symmetric phase) or forming complex conjugate pairs (broken-$\cal PT$ phase); the spontaneous $\cal PT$ breaking hence takes place on the real-$\e$ axis. In the meanwhile, this effective Hamiltonian has the form of Eq.~(\ref{eq:H_ph*}) and hence satisfies non-Hermitian particle-hole symmetry as well, and its spontaneous symmetry breaking takes place on the imaginary-$\e$ axis. Therefore, the spontaneous breaking (restoration) of these two symmetries can take place \textit{simultaneously} at $\e=0$.

This finding is consistent with the fact that the unitary operator $\cal C$ (now given by $\sigma_z$) anticommutes with $\cal P$, and consequently $\cal CT$ and $\cal PT$ can share the same eigenstates and exceptional point. Note however, the symmetric phases of these two symmetries are separated from each other by $\e=0$: it is on the imaginary-$\e$ axis for non-Hermitian particle-hole symmetry and on the real-$\e$ axis for $\cal PT$ symmetry. Therefore, the aforementioned simultaneity indicates that when one of the symmetries is spontaneously broken for a pair of eigenstates, the other symmetry is spontaneously restored. 

\begin{figure}[b]
\includegraphics[clip,width=\linewidth]{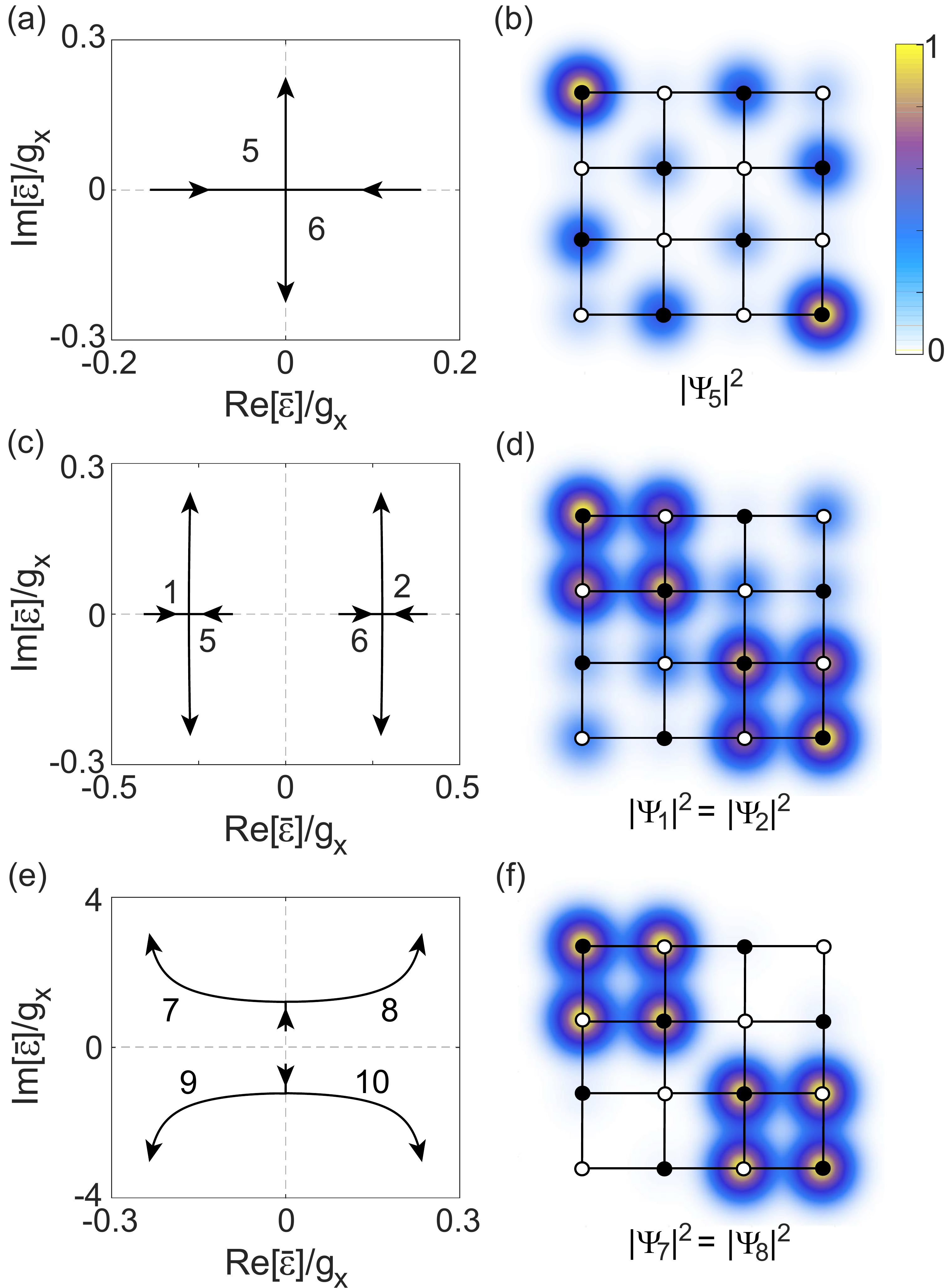}
\caption{(a) Simultaneous restoration of non-Hermitian particle-hole symmetry and breaking of $\cal PT$ symmetry for mode 5 and 6 in Fig.~\ref{fig:ph*}(a). The pump is applied to the entire sublattice $A$ [filled dots in (b)], and the arrows indicate the directions of increasing pump. The spatial profile of mode 5 at the laser threshold ($\gamma_A\approx0.55g_x$) is shown in (b). (c) $\cal PT$ symmetry breaking for mode 1, 2, 5 and 6 in Fig.~\ref{fig:ph*}(a) without restoring their particle-hole symmetry. The pump is applied to the eight cavities in the two $2\times2$ diagonal blocks. The identical spatial profile of mode 1, 2 at the laser threshold ($\gamma_A=\gamma_B\approx0.59g_x$) is shown in (d). (e) Breaking of non-Hermitian particle-hole symmetry for two pairs of modes without restoring their $\cal PT$ symmetry. The pump profile is the same as in (c) and $\gamma_A=\gamma_B\in[6.7~13.4]g_x$. The identical spatial profile of mode 7, 8 at the highest pump strength is shown in (f).}\label{fig:ph*_breaking2}
\end{figure}

This property holds beyond a simple PM. For example, by pumping the entire sublattice $A$, the rectangular lattice in Fig.~\ref{fig:ph*} also has both non-Hermitian particle-hole symmetry and $\cal PT$ symmetry, which apply to the shifted effective Hamiltonian $\bar{H}=\tilde{H}+i(\kappa_0-\gamma_A/2)$ \cite{EP_CMT}. This transform from $\tilde{H}$ to $\bar{H}$ leaves the eigenstates invariant, and the corresponding eigenvalues are simply shifted to $\bar{\e}_n=\e_n+i(\kappa_0-\gamma_A/2)$. The simultaneous breaking (restoration) of non-Hermitian particle-hole symmetry and $\cal PT$ symmetry at $\bar{\e}=0$ is shown in Fig.~\ref{fig:ph*_breaking2}(a).
It can be shown that the spatial overlap between this pump profile and any two passive modes, denoted by $\langle m|A|n\rangle$, is nonzero only when the two modes satisfy $\Psi_m = {\cal CT}\Psi_n$: it is given by $|\langle m|A|n\rangle|=0.5$ when $\langle m|m\rangle=\langle n|n\rangle=1$. As a result, only such particle-hole symmetric partners are strongly coupled at a weak pump strength, and they undergo a spontaneous restoration of particle-hole symmetry and break $\cal PT$ symmetry at the same time. The latter is reflected by the lack of parity symmetry about both the vertical and horizontal axes in the intensity profile shown in Fig.~\ref{fig:ph*_breaking2}(b); note that the system does have $\pi$-rotation symmetry with this pump configuration.

Spontaneous symmetry restoration and breaking of particle-hole and $\cal PT$ symmetries however, does not take place simultaneously in general \cite{Malzard}. For our selectively pumped photonic lattice laser, one example is shown in Fig.~\ref{fig:ph*_breaking2}(c) where we pump the eight cavities in the two $2\times2$ blocks on the diagonal of the rectangular lattice: two pairs of eigenstates undergo $\cal PT$ breaking away from the imaginary axis, i.e., without entering the symmetric phase of non-Hermitian particle-hole symmetry simultaneously. For a spontaneous breaking of non-Hermitian particle-hole symmetry to take place without restoring $\cal PT$ symmetry, it needs to occur away from the real axis, which we illustrate, just for a proof of principle, by going above the laser threshold with the same pump profile in Fig.~\ref{fig:ph*_breaking2}(c) and ignoring any nonlinear effects. The results are shown in Figs.~\ref{fig:ph*_breaking2}(e) and (f), where the intensity profiles of the eigenstates resemble closely the pump profile due to gain guiding \cite{Siegman} with a strong pump strength.

\section{Conclusion and Discussion}

In summary, we have discussed the consequences of chiral symmetry and partile-hole symmetry in non-Hermitian systems. Although both of them lead to a symmetric eigenvalue spectrum $\e_m=-\e_n (m\neq n)$ in a Hermitian system, this property holds only for the chiral symmetry once the system is non-Hermitian. The non-Hermitian particle-hole symmetry requires $\e_m=-\e_n^* (m\neq n)$ instead in its broken-symmetry phase, and its symmetric phase now extends along the imaginary-$\e$ axis and features eigenstates with a unique phase distribution.

Using the zero modes protected by non-Hermitian particle-hole symmetry in a photonic lattice and the spatial degrees of freedom they offer, we have proposed a single-mode, fixed-frequency, and spatially tunable zero-mode laser. Note that the system does not need to have zero modes before the localized pump is applied, which is illustrated by a rectangular lattice with different coupling coefficients $g_x$, $g_y$ in the horizontal and vertical directions; zero modes are created by spontaneous restoration of particle-hole symmetry, and by modifying this process using different pump configurations, we have demonstrated a versatile way to tune the spatial profile of our zero-mode laser, with its lasing frequency pinned at the zero energy, i.e., the single-cavity frequency $\omega_0$ at or closest to the center frequency of the gain curve. Even when $\omega_0$ varies from cavity to cavity and hence lifts the particle-hole symmetry, we have shown that the frequency variation of our zero-mode laser is comparable or even smaller than a single-cavity laser. We note that the variation of cavity decay rate $\kappa_0$ does not lift the particle-hole symmetry, nor does the position and coupling disorder in the photonic lattice.

We have shown that particle-hole symmetry is a general property of non-Hermitian photonic systems consisting of two sublattices. As a result, it exists in $\cal PT$-symmetric lasers demonstrated previously, including the photonic molecule lasers \cite{Hodaei,EP_exp,Yang}. There is another symmetry in non-Hermitian system that is related to both particle-hole symmetry and $\cal PT$ symmetry, which was termed antisymmetric $\cal PT$ symmetry (or anti-$\cal PT$ symmetry for short) \cite{antiPT}, where the effective Hamiltonian anticommutes with a combined parity and time reversal operation. In fact, it can be treated as a special case of non-Hermitian particle-hole symmetry, where the general unitary operator $\cal C$ takes the form of the parity operator. Anti-$\cal PT$ symmetry and its spontaneous symmetry breaking have been observed in an atomic system, where two beams of flying atoms are coupled via coherent transport \cite{antiPT_exp}.

Although here we do not carry out the nonlinear calculation when the pump strength is above the laser threshold, the discussions we have presented, including those on the single-mode performance of our zero-mode laser, should hold close to the laser threshold where the nonlinear effect is weak. This approach has been verified for $\cal PT$-symmetric lasers by comparing with the steady-state solutions of the Maxwell-Bloch equations, obtained by the SALT formulism \cite{EP_CMT,Nonl_PT}. Since neither gain saturation nor loss saturation lifts non-Hermitian particle-hole symmetry of $\tilde{H}$ given by Eq.~(\ref{eq:H_ph*}), it is possible that more than one zero mode can lase simultaneously when the pump strength is high above the first laser threshold. This scenario of symmetry-enforced degenerate lasing is beyond our current discussion and will be studied in a future work. Nevertheless, it is important to note that these saturation effects prevent a huge contrast of gain and loss, which could otherwise lead to a lasing frequency deviated from the zero energy via spontaneous breaking of particle-hole symmetry at a high pump strength [see Fig.~\ref{fig:ph*_breaking2}(e)].

Finally, we comment that while a spatially tunable laser with a fixed frequency can be achieved by other means, for example, using a spatial light modulator \cite{SLM} after a single-mode laser or a degenerate cavity laser with a variable aperture \cite{DCL1,DCL2}, they require conventional optical elements and hence cannot be easily integrated on a single chip. Our zero-mode photonic lattice laser, in contrast, does not suffer from this constraint, especially when the cavities are connected to individual electrodes and pumped electrically. Such a zero-mode laser may find applications in telecommunication, where spatial encoding is held by some to be last frontier of signal processing.

L.G. acknowledges support by the NSF under Grant No. DMR-1506987.


\begin{thebibliography}{99}



\bibitem{Hasan} M. Z. Hasan and C. L. Kane, \textit{Topological insulators}, Rev. Mod. Phys. \textbf{82}, 3045 (2010).
\bibitem{Qi} X.-L. Qi and S.-C. Zhang, \textit{Topological insulators and superconductors}, Rev. Mod. Phys. \textbf{83}, 1057 (2011).
\bibitem{Alicea} J. Alicea, \textit{New directions in the pursuit of Majorana fermions in solid state systems}, Rep. Prog. Phys. \textbf{75}, 076501 (2012).
\bibitem{Beenakker} C. W. J. Beenakker, \textit{Random-matrix theory of Majorana fermions and topological superconductors}, Rev. Mod. Phys. \textbf{87},  1037 (2015).

\bibitem{Sarma_RMP} C. Nayak, S. H. Simon, A. Stern, M. Freedman, and S. Das Sarma, \textit{Non-Abelian anyons and topological quantum computation}, Rev. Mod. Phys. \textbf{80}, 1083 (2008).
\bibitem{Sarma_QI} S. D. Sarma, M. Freedman, and C. Nayak, \textit{Majorana zero modes and topological quantum computation}, npj Quantum Information \textbf{1}, 15001 (2015).
\bibitem{Alicea_PRX} D. Aasen, M. Hell, R. V. Mishmash, A. Higginbotham, J. Danon, M. Leijnse, T. S. Jespersen, J. A. Folk, C. M. Marcus, K. Flensberg, and J. Alicea \textit{Milestones toward Majorana-based quantum computing}, Phys. Rev. X \textbf{6}, 031016 (2016).

\bibitem{Lu} L. Lu, J. D. Joannopoulos, and M. Soljacic, \textit{Topological photonics}, Nat. Photon. \textbf{8}, 821 (2014).

\bibitem{Malzard} S. Malzard, C. Poli, and H. Schomerus, \textit{Topologically protected defect states in open photonic systems with non-Hermitian charge-conjugation and parity-time symmetry}, Phys. Rev. Lett. \textbf{115}, 200402 (2015).
\bibitem{EP_CMT} R.~El-Ganainy, M.~Khajavikhan, and L.~Ge, \textit{Exceptional points and lasing self-termination in photonic molecules}, Phys.~Rev.~A \textbf{90}, 013802 (2014).

\bibitem{Pikulin} D. I. Pikulin and Y. V. Nazarov, \textit{Topological properties of superconducting junctions}, JETP Lett. \textbf{94}, 693 (2012).
\bibitem{Ryu} S. Ryu and Y. Hatsugai, \textit{Topological origin of zero-energy edge states in particle-hole symmetric systems}, Phys. Rev. Lett. \text{89}, 077002 (2002).


\bibitem{Rechtsman} M. C. Rechtsman, J. M. Zeuner, A. Tünnermann, S. Nolte, M. Segev, and A. Szameit, \textit{Strain-induced pseudomagnetic field and photonic Landau levels in dielectric structures}, Nat. Photon. \textbf{153} (2012).
\bibitem{Vicencio} R. A. Vicencio, C. Cantillano, L. Morales-Inostroza, B. Real, C. Mej\'ia-Cort\'es, S. Weimann, A. Szameit, and M. I. Molina,
\textit{Observation of localized states in Lieb photonic lattices}, Phys. Rev. Lett. \textbf{114}, 245503 (2015).
\bibitem{Mukherjee} S. Mukherjee, A. Spracklen, D. Choudhury, N. Goldman, P. \"Ohberg, E. Andersson, and R. R. Thomson,
\textit{Observation of a localized flat-band state in a photonic Lieb lattice}, Phys. Rev. Lett. \textbf{114}, 245504 (2015).
\bibitem{Baboux} F. Baboux, L. Ge, T. Jacqmin, M. Biondi, E. Galopin, A. Lema\^{i}tre, L. Le Gratiet, I. Sagnes, S. Schmidt, H. E. T\"ureci, A. Amo, and J. Bloch \textit{Bosonic condensation and disorder-induced localization in a flat band}, Phys. Rev. Lett. \textbf{116}, 066402 (2016).
\bibitem{Schomerus} C. Poli, M. Bellec, U. Kuhl, F. Mortessagne, and H. Schomerus, \textit{Selective enhancement of topologically induced interface states in a dielectric resonator chain}, Nat. Commun. \textbf{6}, 6710 (2015).




\bibitem{Bender1} C.~M.~Bender and S.~Boettcher, \textit{Real spectra in non-Hermitian hamiltonians having $\cal PT$ symmetry},
Phys. Rev. Lett. {\bf 80}, 5243 (1998).

\bibitem{El-Ganainy_OL06} R.~El-Ganainy, K.~G.~Makris, D.~N.~Christodoulides, and Z.~H.~Musslimani,
\textit{Theory of coupled optical $\cal PT$-symmetric structures},
Opt. Lett. {\bf 32}, 2632 (2007).

\bibitem{Moiseyev} S.~Klaiman, U.~Gunther, and N.~Moiseyev,
\textit{Visualization of branch points in $\cal PT$-symmetric waveguides},
Phys. Rev. Lett. {\bf 101}, 080402 (2008).

\bibitem{Makris_prl08} K.~G.~Makris, R.~El-Ganainy, D.~N.~Christodoulides, and Z.~H.~Musslimani,
\textit{Beam dynamics in $\cal PT$ symmetric optical lattices},
Phys. Rev. Lett. {\bf 100}, 103904 (2008).

\bibitem{Longhi} S.~Longhi,
\textit{$\cal PT$-symmetric laser absorber},
Phys. Rev. A {\bf 82}, 031801(R) (2010).

\bibitem{CPALaser} Y.~D.~Chong, L.~Ge, and A.~D.~Stone,
\textit{$\cal PT$-symmetry breaking and laser-absorber modes in optical scattering systems},
Phys. Rev. Lett. {\bf 106}, 093902 (2011).


\bibitem{RC}Z.~Lin, J.~Schindler, F.~M.~Ellis, and T.~Kottos,
\textit{Unidirectional invisibility induced by $\cal PT$-symmetric periodic structures},
Phys. Rev. A {\bf 85}, 050101(R) (2012).

\bibitem{Microwave} S. Bittner, B. Dietz, U. G\"unther, H. L. Harney, M.
Miski-Oglu, A. Richter, and F. Sch\"afer,
\textit{PT symmetry and spontaneous symmetry breaking in a microwave billiard},
Phys. Rev. Lett. {\bf 108}, 024101 (2012).

\bibitem{Regensburger} A. Regensburger, C. Bersch, M.-A. Miri, G. Onishchukov, D. N. Christodoulides, U. Peschel,
\textit{Parity-time synthetic photonic lattices}, Nature (London) {\bf 488}, 167 (2012).

\bibitem{PTConservation} L.~Ge, Y.~D.~Chong, and A. D. Stone,
\textit{Conservation Relations and Anisotropic Transmission Resonances in One-Dimensional $\cal PT$-Symmetric Photonic Heterostructures},
Phys. Rev. A {\bf 85}, 023802 (2012).
\bibitem{Robin} P.~Ambichl, K.~G.~Makris, L.~Ge, Y.~D.~Chong, A.~D.~Stone, and S.~Rotter,
\textit{Breaking of $\cal PT$ symmetry in bounded and unbounded scattering systems},
Phys. Rev. X {\bf 3}, 041030 (2013).
\bibitem{Ge_PRX} L.~Ge and A. D. Stone, \textit{Parity-time symmetry breaking beyond one dimension: the role of degeneracy},
Phys. Rev. X {\bf 4}, 031011 (2014).
\bibitem{Feng} L. Feng, Z. J.Wong, R.-M.Ma, Y.Wang, and X. Zhang, \textit{Singlemode laser by parity-time symmetry breaking}, Science \textbf{346}, 972 (2014).
\bibitem{Hodaei} H. Hodaei, M.-A. Miri, M. Heinrich, D. N. Christodoulides, and M. Khajavikhan, \textit{Parity-time-symmetric microring lasers}, Science \textbf{346}, 975 (2014).
\bibitem{Yang} B. Peng, S. K. \"Ozdemir, F. Lei, F. Monifi, M. Gianfreda, G. L. Long, S. Fan, F. Nori, C. M. Bender, and L. Yang
\textit{Parity-time-symmetric whispering-gallery microcavities}, Nat. Phys. \textit{10}, 394 (2014).

\bibitem{PT_RMP} V. V. Konotop, J. Yang, and D. A. Zezyulin, \textit{Nonlinear waves in $\mathcal{PT}$-symmetric systems}, Rev. Mod. Phys. \textbf{88}, 35002 (2016).

\bibitem{SPASALT} L.~Ge, Y.~D.~Chong, and A.~D.~Stone, \textit{Steady-state ab initio laser theory: generalizations and analytic results}, Phys. Rev. A \textbf{82}, 063824 (2010).

\bibitem{LiebThm} E. H. Lieb, \textit{Two theorems on the Hubbard model}, Phys. Rev. Lett. \textbf{62}, 1201 (1989).
\bibitem{Wilczek} F. Wilczek, \textit{Fractional statistics and anyon superconductivity} (World Scientific, Singapore, 1990).




\bibitem{EP7} M.~Liertzer, L.~Ge, A.~Cerjan, A.~D.~Stone, H.~E.~T\"{u}reci, and S.~Rotter,
\textit{Pump-induced exceptional points in lasers},
Phys. Rev. Lett. {\bf 108}, 173901 (2012).
\bibitem{EP_exp} M. Brandstetter, M. Liertzer, C. Deutsch, P. Klang, J. Sch\"oberl, H. E. T\"ureci, G. Strasser, K. Unterrainer and S. Rotter,
\textit{Reversing the pump dependence of a laser at an exceptional point}, Nat. Comm. {\bf 5}, 4034 (2014).
\bibitem{Nonl_PT} L. Ge and R. El-Ganainy, \textit{Nonlinear modal interactions in parity-time (PT) symmetric lasers}, Sci. Rep. \textbf{6}, 24889 (2016).

\bibitem{Narimanov} S. Shinohara, T. Harayama, T. Fukushima, M. Hentschel, T. Sasaki, and E. E. Narimanov, \textit{Chaos-assisted directional light emission from microcavity lasers}, Phys. Rev. Lett. \textbf{104}, 163902 (2010).


\bibitem{EP1} J. Okolowicz, M. Ploszajczak, and I. Rotter,
\textit{Dynamics of quantum systems embedded in a continuum},
Phys. Rep. {\bf 374}, 271 (2003).

\bibitem{EP2} W. D. Heiss, \textit{Exceptional points of non-Hermitian operators},
J. Phys. A: Math. Gen. {\bf 37}, 2455 (2004).

\bibitem{EPMVB} M. V. Berry, \textit{Physics of nonhermitian degeneracies},
Czechoslovak J. Phys. {\bf 54}, 1039 (2004).

\bibitem{EP3} N. Moiseyev, {\it Non-Hermitian Quantum Mechanics} (Cambridge, New York, 2011).

\bibitem{EP4} C. Dembowski, H.-D. Gr\"af, H. Harney, A. Heine, W. Heiss, H. Rehfeld, and A. Richter,
\textit{Experimental observation of the topological structure of exceptional points},
Phys. Rev. Lett. {\bf 86}, 787 (2001).

\bibitem{EP5} J. Wiersig, S.-W. Kim, and M. Hentschel,
\textit{Asymmetric scattering and nonorthogonal mode patterns in optical microspirals},
Phys. Rev. A {\bf 78}, 053809 (2008).

\bibitem{EP6} S.-B. Lee, J. Yang, S. Moon, S.-Y. Lee, J.-B. Shim, S. Kim, J.-H. Lee, and K. An,
\textit{Observation of an exceptional point in a chaotic optical microcavity},
Phys. Rev. Lett. {\bf 103}, 134101 (2009).

\bibitem{EP8} L. Ge, Y. D. Chong, S. Rotter, H. E. T\"{u}reci, and A. D. Stone, \textit{Unconventional modes in lasers with spatially varying gain and loss}, Phys. Rev. A {\bf 84}, 023820 (2011).

\bibitem{Tureci_PRA} H.~E.~T\"{u}reci, A.~D.~Stone, and L.~Ge, \textit{Theory of the spatial structure of nonlinear lasing modes}, Phys. Rev. A \textbf{76}, 013813 (2007).

\bibitem{antiPT} L. Ge and H. E. T\"ureci,
\textit{Antisymmetric PT-photonic structures with balanced positive- and negative-index materials},
Phys. Rev. A {\bf 88}, 053810 (2013).
\bibitem{antiPT_exp} P. Peng, W. Cao, C. Shen, W. Qu, J. Wen, L. Jiang, and Y. Xiao, 
\textit{Anti-parity-time symmetric optics via flying atoms}, Nat. Phys. \textbf{3842} (2016).

\bibitem{Siegman} A. E. Siegman, \textit{Gain-guided, index-antiguided fiber lasers}, J. Opt. Soc. Am. B \textbf{24}, 1677 (2007).
\bibitem{SLM}  M. Leonetti and C. L\'opez, \textit{Active subnanometer spectral control of a random laser}, Appl. Phys. Lett. \textbf{102}, 071105 (2013).
\bibitem{DCL1} M. Nixon, B. Redding, A. A. Friesem, H. Cao, and N. Davidson, \textit{Efficient method for controlling the spatial coherence of a laser}, Opt. Lett. \textbf{38}, 3858 (2013).
\bibitem{DCL2} S. Knitter, C. Liu, B. Redding, M. K. Khokha, M. A. Choma, and H. Cao, \textit{Coherence switching of a degenerate VECSEL for multimodality imaging}, Optica \textbf{3}, 403 (2016).
    
    



\end{thebibliography}
\end{document}